\documentclass[conference]{IEEEtran}
\makeatletter
\def\ps@headings{%
\def\@oddhead{\mbox{}\scriptsize\rightmark \hfil \thepage}%
\def\@evenhead{\scriptsize\thepage \hfil \leftmark\mbox{}}%
\def\@oddfoot{}%
\def\@evenfoot{}}
\makeatother
\usepackage{amsmath,amsfonts}
\usepackage{amsfonts,amssymb}
\usepackage{algorithmic}
\usepackage[ruled,linesnumbered,vlined]{algorithm2e}
\usepackage{array}
\usepackage{xcolor}
\usepackage{textcomp}
\usepackage{stfloats}
\usepackage{url}
\usepackage{verbatim}
\usepackage{graphicx}
\usepackage{cite}
\usepackage{bm}
\usepackage{float}
\usepackage{subfigure}

\begin{document}

\title{Event-Triggered GAT-LSTM Framework for Attack Detection in Heating, Ventilation, and Air Conditioning Systems}
\author{Zhenan Feng and Ehsan Nekouei* 
\thanks{Zhenan Feng and Ehsan Nekouei are with the Department of Electrical Engineering, City University of Hong Kong. E-mail: {\tt zhenafeng2-c@my.cityu.edu.hk, enekouei@cityu.edu.hk}. }}



\maketitle

\begin{abstract}
Heating, Ventilation, and Air Conditioning (HVAC) systems are essential for maintaining indoor environmental quality, but their interconnected nature and reliance on sensor networks make them vulnerable to cyber-physical attacks. Such attacks can interrupt system operations and risk leaking sensitive personal information through measurement data. In this paper, we propose a novel attack detection framework for HVAC systems, integrating an Event-Triggering Unit (ETU) for local monitoring and a cloud-based classification system using the Graph Attention Network (GAT) and the Long Short-Term Memory (LSTM) network. The ETU performs a binary classification to identify potential anomalies and selectively triggers encrypted data transmission to the cloud, significantly reducing communication cost. The cloud-side GAT module models the spatial relationships among HVAC components, while the LSTM module captures temporal dependencies across encrypted state sequences to classify the attack type. Our approach is evaluated on datasets that simulate diverse attack scenarios. Compared to GAT-only (94.2\% accuracy) and LSTM-only (91.5\%) ablations, our full GAT-LSTM model achieves 98.8\% overall detection accuracy and reduces data transmission to 15\%. These results demonstrate that the proposed framework achieves high detection accuracy while preserving data privacy by using the spatial-temporal characteristics of HVAC systems and minimizing transmission costs through event-triggered communication.  
\end{abstract}


\section{Introduction}
\subsection{Motivation}
Heating, Ventilation, and Air Conditioning (HVAC) systems are integral to modern buildings, ensuring occupant comfort and safety by regulating temperature, humidity, and air quality. These systems rely on interconnected components and extensive sensor networks to monitor and adjust environmental conditions. However, this reliance on networked sensors introduces significant vulnerabilities, exposing HVAC systems to cyber-physical attacks. Such attacks can manipulate sensor measurements, disrupt system functionality, or lead to unauthorized access to sensitive personal information embedded in the data \cite{attackresult1}.

The challenges of protecting HVAC systems and safeguarding user privacy have not been effectively addressed. Conventional fault detection and anomaly detection methods, such as rule-based or threshold-based approaches, fail to address the complexity of cyber-physical attacks, which often exploit both spatial relationships among HVAC components and temporal dependencies in system behavior. Furthermore, transmitting raw measurement data to cloud-based services for analysis raises serious privacy concerns, as it risks the leakage of sensitive occupant information \cite{buildingchallenges}.

Graph-based learning has become a powerful approach for analyzing interconnected systems such as HVAC. Graph Attention Networks (GAT) \cite{GAT} capture spatial relationships by assigning greater weight to critical interactions among components. Meanwhile, Long Short-Term Memory (LSTM) networks \cite{LSTM} excel in modeling temporal dependencies, making them ideal for detecting evolving patterns in sequential data \cite{canaan2023lstm}. Despite the effectiveness of these individual techniques, few works have explored a unified approach that leverages both the spatial and temporal dimensions, which is essential for detecting and classifying complex cyber-physical attacks.

To address these challenges, we propose a two-tiered framework combining local fault detection with cloud-based attack classification. On the local side, a lightweight ETU continuously monitors the HVAC system and performs real‑time binary classification to flag anomalies. When the ETU raises an alarm, the ETU triggers a Fully Homomorphic Encryption (FHE) scheme to secure the state data before transmission, greatly reducing communication and preserving occupant privacy. On the cloud side, GAT captures spatial relationships among components, and LSTM networks model temporal dynamics. Together, they accurately classify the type of attack.

By bridging these gaps, the proposed framework not only enhances the security and reliability of HVAC systems, but also provides a scalable, privacy-preserving approach to attack detection. 

\subsection{Related Work}
The interconnected nature of HVAC systems makes them highly vulnerable to cyber-physical attacks. These attacks not only disrupt operational efficiency, but also risk exposing sensitive personal information through sensor data, requiring advanced detection mechanisms. Recent studies \cite{tree,rfhybrid,hvacml}, have explored various methodologies to detect such threats. 
Using techniques such as Decision Trees \cite{DT}, Random Forests (RF) \cite{RF}, and Gradient Boosting, researchers have demonstrated the capability to accurately identify security breaches using sensor data obtained from HVAC systems. Reference \cite{tree} established a decision tree model to diagnose abrupt faults and process faults. Reference \cite{rfhybrid} used RF algorithm to quickly classify normal events and attack events. Reference \cite{hvacml} employed machine learning models to monitor system behaviors and identify deviations indicative of sabotage. Such a model achieved high detection rates, such as 99\% accuracy and 98.2\% recall, showcasing their potential for real-time applications.
Although the above methods can achieve enough accuracy, they cannot utilize the interconnected nature of the system. Thus, when detecting an attack on a complex building, the influence of one part of the HVAC system may be ignored, causing the wrong detection result.

Graph Neural Networks (GNNs) have emerged as a promising solution to model interconnected systems, offering the ability to capture spatial relationships between components \cite{gnn2025survey}. GNNs have been successfully applied in cyber-physical systems, including HVAC and smart grids, to detect anomalies and model spatial dependencies. For example, convolutional GNNs have been used to analyze telemetry data from interconnected devices, demonstrating efficient anomaly detection performance \cite{gnn}. In power systems, GNNs have been utilized to simulate the topological and functional properties of cyber-physical grids, creating realistic benchmarks for evaluating system resilience \cite{gnngrid}. Furthermore, spatio-temporal deep graph networks have been applied in smart grids to jointly model spatial, temporal, and node-level features for detecting and classifying events \cite{gnnspatio}.

Among various GNN variants, GATs provide distinct advantages in modeling spatial dependencies in HVAC systems. Unlike conventional GNNs that rely on fixed aggregation schemes, GATs dynamically assign attention weights to edges, prioritizing critical relationships between nodes \cite{GAT}. This is particularly valuable in HVAC systems, where certain components may significantly influence the overall behavior of the system during an attack. By focusing on these key connections, GATs enhance interpretability and robustness in detecting attack patterns \cite{gatreview}.

Despite these advancements, there is limited research on integrating GAT-based approaches with temporal modeling techniques for HVAC attack detection. HVAC systems display both spatial dependencies, representing the relationships among system components, and temporal dynamics, capturing sequential changes in system states. Addressing both spatial and temporal complexities requires a unified framework using GATs for spatial modeling and LSTM networks for temporal analysis \cite{lstmtraffic}. By integrating these techniques, it becomes possible to effectively model spatial-temporal patterns \cite{gnntemporal}, enhancing the robustness and accuracy of attack detection. Such an approach not only addresses the limitations of traditional machine learning methods but also provides a comprehensive solution for detecting and mitigating cyber-physical attacks in HVAC systems.



\section{The Proposed Attack Detection Framework For HVAC System}
\begin{figure}[h]
    \centering
    \includegraphics[width=3.4in]{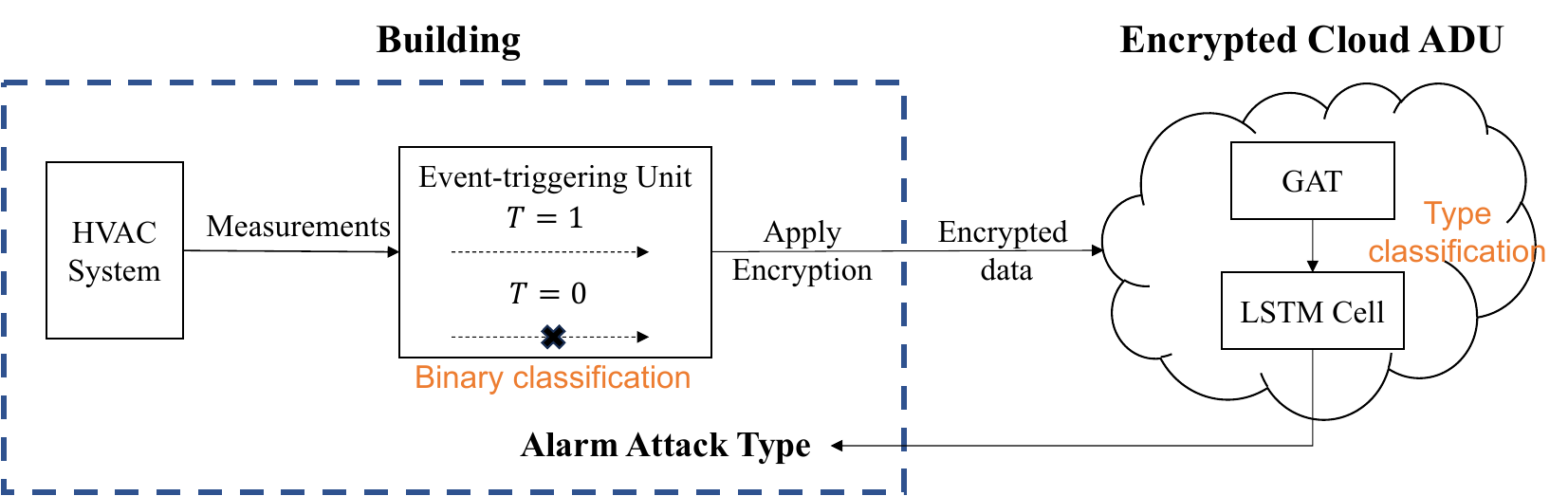}
    \caption{The overview of the proposed framework.}
    \label{fig:framework}
\end{figure}

In this paper, we develop an encrypted HVAC system attack detection framework, as shown in Fig. \ref{fig:framework}. The framework integrates a ETU at the local side and a cloud-based attack type identification unit to efficiently monitor, detect, and classify cyber-physical attacks on HVAC systems. The ETU operates locally to continuously monitor the system's measurements and performs a binary classification to determine whether there is a potential attack. If an anomaly is detected, the ETU encrypts the measurements and transmits them to the cloud-based ADU for detailed type classification.

To preserve the privacy of building occupants, all measurements are encrypted using a Fully Homomorphic Encryption (FHE) scheme prior to transmission. This ensures that sensitive personal information embedded in the measurements is protected throughout the communication process. By using an event-triggered mechanism, the ETU reduces communication overhead by only sending measurements to the cloud when a potential attack is detected. This design optimizes resource usage by minimizing unnecessary communication and computation costs while ensuring timely detection and response to threats.
In the cloud, the ADU processes the encrypted data to classify the type of attack. The ADU combines GAT that models spatial relationships between HVAC components, and LSTM networks to capture temporal dependencies in the data. This architecture enables detailed attack classification while preserving data privacy through FHE. Together, the ETU and ADU provide a scalable and efficient solution for detecting and classifying cyber-physical attacks.

In this section, we first describe the underlying model of the HVAC system, which forms the basis of the framework. Next, we detail the design of the event-triggering unit and its role in monitoring and detecting potential attacks, including the structure of the GAT used in local feature extraction. Finally, we present the cloud-based attack detection unit, including its encrypted processing mechanism and the rationale for employing LSTM networks to classify attack types based on encrypted measurements.

\subsection{HVAC System Model}

The HVAC system in a building is modeled as a network of interconnected components, such as fans, condensers, and other devices, responsible for the regulation of indoor environmental conditions. Each room in the building is served by a HVAC subsystem, which is represented as a separate graph $G = (V, E)$, where $V$ denotes the set of nodes corresponding to the components in the room and $E$ denotes the set of edges representing physical or logical connections between the components.

For each graph, each node $ v \in V $ is associated with a dynamic feature vector which captures its measurements (\emph{e.g.}, temperature) at time $ t $. The adjacency matrix $ A $ defines the relationships between nodes, with edge weights encoding contextual information such as connection strength. These individual graph representations enable both local monitoring by the ETU and global attack detection by the ADU. By modeling each room's HVAC system as an independent graph, the framework can efficiently analyze spatial and temporal characteristics within each room while maintaining scalability for larger buildings.
\subsection{Event-Triggering Unit (ETU)}

The ETU operates locally to continuously monitor the state of the HVAC system and minimize communication cost. Its primary goals are to detect potential attacks using a lightweight neural network and selectively transmit encrypted data to the cloud-based ADU for further analysis. The ETU achieves this through the following mechanisms:

\paragraph{Neural Network for Binary Classification}
The ETU employs a multilayer perceptron (MLP) to classify the state of the HVAC system at each time step $ t $. The network processes the measurement vector $ \textit{\textbf{x}}_t $ and determines whether the system is operating normally ($ y_t^{\text{local}} = 0 $) or shows signs of a potential attack ($ y_t^{\text{local}} = 1 $). The network is defined as:
\begin{align}
    y_t^{\text{local}} &= \arg\max ( \text{softmax}( \textit{\textbf{W}}^3 \sigma_2( \textit{\textbf{W}}^2 \sigma_1(\textit{\textbf{W}}^1 \textit{\textbf{x}}_t + \textit{\textbf{b}}^1 ) \nonumber \\ &+ \textit{\textbf{b}}^2 ) + \textit{\textbf{b}}^3 )),
\end{align}
where $ \textit{\textbf{W}}^1, \textit{\textbf{W}}^2, \textit{\textbf{W}}^3 $ and $ \textit{\textbf{b}}^1, \textit{\textbf{b}}^2, \textit{\textbf{b}}^3 $ are the weights and biases of the input, hidden, and output layers, respectively. The mappings $ \sigma^1(\cdot)$ and $\sigma^2(\cdot) $ are non-linear activation functions (\emph{e.g.}, ReLU). The output layer applies a softmax activation to predict the probability of each class (normal or potential attack).

The MLP enable the ETU to capture complex patterns in the measurement data while maintaining computational efficiency for local processing.

\paragraph{Triggering Mechanism}  
If the neural network detects a potential attack ($ y_t^{\text{local}} = 1 $), the ETU computes a deviation score $ f(\textit{\textbf{x}}_t) $ to evaluate the severity of the anomaly. The deviation score quantifies how much the current measurements $ \textit{\textbf{x}}_t $ deviate from the expected normal behavior of the system. It is computed as:
\begin{equation}
    f(\textit{\textbf{x}}_t) = P(y_t^{\text{local}} = 1),
\end{equation}
where $ P(y_t^{\text{local}} = 1) $ represents the probability assigned by the neural network to the anomalous class. This score reflects the confidence of the ETU in detecting a potential attack. 

Data is transmitted to the cloud-based ADU only if the deviation score exceeds a predefined threshold $ \tau $:
\begin{equation}
    \mathcal{T}(\textit{\textbf{x}}_t) = 
    \begin{cases} 
    1, & \text{if } f(\textit{\textbf{x}}_t) > \tau, \\
    0, & \text{otherwise}.
    \end{cases}
\end{equation}
This mechanism reduces unnecessary transmissions by ensuring that only significant anomalies are sent to the cloud for further analysis.

When the triggering condition $ \mathcal{T}(\textit{\textbf{x}}_t) = 1 $ is met, a FHE scheme is applied to measurements:
\begin{equation}
    D_t = \text{Encrypt}(\textit{\textbf{x}}_t).
\end{equation}
This ensures that sensitive personal information embedded in the measurements remains secure during communication with the cloud. The process of ETU is shown in Algorithm~\ref{alg:etu}. Note that ETU sends a sequence of data points to the cloud for further type classification.

\begin{algorithm}[h]
\caption{ETU and Encryption Operations}
\label{alg:etu}
\KwIn{Measurements $ \textit{\textbf{x}}_t $, threshold $ \tau $, and encryption scheme.}
\KwOut{Encrypted data $ D_t $ or no transmission.}
\For{each time step $ t $}{
    Compute $ y_t^{\text{local}} \leftarrow \arg\max \text{softmax}(\textit{\textbf{W}}^3 \sigma^2(\textit{\textbf{W}}^2 \sigma^1(\textit{\textbf{W}}^1 \textit{\textbf{x}}_t + \textit{\textbf{b}}^1) + \textit{\textbf{b}}^2) + \textit{\textbf{b}}^3) $\;
    \If{$ y_t^{\text{local}} = 1 $}{
        Compute $ \mathcal{T}(\textit{\textbf{x}}_t) \leftarrow f(\textit{\textbf{x}}_t) $\;
        \If{$ \mathcal{T}(\textit{\textbf{x}}_t) > \tau $}{
            Encrypt $ D_t \leftarrow \text{Encrypt}(\textit{\textbf{x}}_t) $\;
            Transmit $ D_t $\;
        }
    }
}
\end{algorithm}

\subsection{Cloud-Based ADU}

The ADU resides in the cloud and aims at classifying the type of attack when a potential anomaly is detected by the ETU. It leverages advanced graph-based and sequence-based models to extract spatial and temporal features from the encrypted data.

\paragraph{GAT for Spatial Analysis}  

In this graph, each node $ v \in V $ corresponds to an HVAC component (\emph{e.g.}, supply fan, condenser) and is associated with a feature vector $\textit{\textbf{h}}_i$. The GAT dynamically assigns attention weights to these connections, enabling the model to prioritize the most critical relationships for accurate analysis.

To update node features, the GAT computes attention coefficients for each pair of connected nodes based on their feature similarity:
\begin{equation}
    \alpha_{ij} = \frac{\text{exp}\Big( \text{LeakyReLU}\big(\textit{\textbf{a}}^\top [ \textit{\textbf{W}}_{\alpha} \textit{\textbf{h}}_i \,||\, \textit{\textbf{W}}_{\alpha} \textit{\textbf{h}}_j ]\big) \Big)}{\sum_{k\in\mathcal{N}(i)}\text{exp}\Big( \text{LeakyReLU}\big(\textit{\textbf{a}}^\top [ \textit{\textbf{W}}_{\alpha} \textit{\textbf{h}}_i \,||\, \textit{\textbf{W}}_{\alpha} \textit{\textbf{h}}_k ]\big) \Big)},
\end{equation}
where $ \mathcal{N}(i) $ denotes the neighbors of node $ i $, $\textit{\textbf{W}}_{\alpha}$ is a learnable weight matrix, $\textit{\textbf{a}}$ is the scaling vector, and $||$ denotes concatenation. The updated feature $\mathbf{h}_i$ for node $i$ is then given by:
\begin{equation}
    \mathbf{h}_i = \sigma\left( \sum_{j \in \mathcal{N}(i)} \alpha_{ij} \textit{\textbf{W}}_{\beta} \textit{\textbf{h}}_j \right),
\end{equation}
where $\textit{\textbf{W}}_{\beta}$ is a scaling matrix and $ \sigma(\cdot) $ is a non-linear activation function such as ReLU.

The graph embedding $ \textit{\textbf{g}}_t $ is computed by aggregating the updated features $ \mathbf{h}_i $ of all nodes in the graph at time $ t $ (\emph{e.g.}, through mean pooling). These embeddings serve as the input sequence for the LSTM described in the next subsection, enabling robust spatial and temporal analysis.

\paragraph{LSTM for Temporal Modeling}  
The temporal dependencies in the sequence of graph embeddings are modeled using a LSTM network. The graph embeddings $ \{\textit{\textbf{g}}_{t-T+1}, \dots, \textit{\textbf{g}}_t\} $, computed by aggregating the spatial features $ \mathbf{h}_i $ from the GAT, represent the system’s spatial states over a time window. These embeddings are processed by the LSTM to capture temporal patterns critical for classifying attack types.
\addtolength{\topmargin}{0.01in}

At each time step $ t $, the LSTM updates its internal states using the current graph embedding $ \textit{\textbf{g}}_t $ and the hidden state $ \textit{\textbf{s}}_{t-1} $ from the previous time step:
\begin{align}
    \textit{\textbf{f}}_t &= \sigma(\textit{\textbf{W}}_\textit{\textbf{f}} [\textit{\textbf{s}}_{t-1}, \textit{\textbf{g}}_t] + \textit{\textbf{b}}_\textit{\textbf{f}}), \\
    \textit{\textbf{e}}_t &= \sigma(\textit{\textbf{W}}_\textit{\textbf{e}} [\textit{\textbf{s}}_{t-1}, \textit{\textbf{g}}_t] + \textit{\textbf{b}}_\textit{\textbf{e}}), \\
    \tilde{\textit{\textbf{C}}}_t &= \tanh(\textit{\textbf{W}}_\textit{\textbf{C}} [\textit{\textbf{s}}_{t-1}, \textit{\textbf{g}}_t] + \textit{\textbf{b}}_\textit{\textbf{C}}), \\
    \textit{\textbf{C}}_t &= \textit{\textbf{f}}_t \odot \textit{\textbf{C}}_{t-1} + \textit{\textbf{e}}_t \odot \tilde{\textit{\textbf{C}}}_t, \\
    \textit{\textbf{s}}_t &= \textit{\textbf{o}}_t \odot \tanh(\textit{\textbf{C}}_t),
\end{align}
where $ \textit{\textbf{s}}_t $ is the hidden state summarizing the temporal context and $ \textit{\textbf{C}}_t $ is the cell state storing long-term information. The gates $ \textit{\textbf{f}}_t $, $ \textit{\textbf{e}}_t $, and $ \textit{\textbf{o}}_t $ control how much information is retained, added, or output, respectively. The functions $ \sigma(\cdot) $ and $ \tanh(\cdot) $ are the sigmoid and hyperbolic tangent activation functions, and $ \odot $ denotes element-wise multiplication.

The final hidden state $ \textit{\textbf{s}}_T $, which summarizes the entire sequence, is passed to the classifier to determine the type of attack. This integration of spatial and temporal modeling enables robust detection by using both static relationships among components and dynamic changes over time.

\paragraph{Attack Type Classification}  
The final hidden state $ \textit{\textbf{s}}_T $, generated by the LSTM after processing the sequence of graph embeddings $ \{\textit{\textbf{g}}_{t-T+1}, \dots, \textit{\textbf{g}}_t\} $, contains both the spatial relationships among HVAC components (captured by the GAT) and the temporal dependencies over the observed time window. This hidden state is used to classify the detected anomaly into one of $ n $ attack types or as normal operation:
\begin{equation}
    y_t^{\text{cloud}} = \arg\max \left( \text{softmax}(\textit{\textbf{W}}_{\text{ADU}} \textit{\textbf{s}}_T + \textit{\textbf{b}}_{\text{ADU}}) \right),
\end{equation}
where $ \textit{\textbf{W}}_{\text{ADU}} $ and $ \textit{\textbf{b}}_{\text{ADU}} $ are the weight matrix and bias vector of the classification layer in the ADU, and $ \text{softmax}(\cdot) $ produces a probability distribution over the possible classes.

The process of attack type classification by the ADU is summarized in Algorithm~\ref{alg:adu}. All computations in the ADU are performed on encrypted data using an FHE scheme, ensuring that sensitive information remains protected during graph-based feature extraction and sequence-based analysis. The FHE scheme enables secure operations to be performed directly on encrypted inputs. Note that activation functions and other non-linear operations are approximated using polynomial functions to ensure compatibility with the FHE.

\begin{algorithm}[h]
\caption{Cloud-Based ADU Operation}
\label{alg:adu}
\KwIn{Encrypted data $ D_t $ and sequence length $ T $.}
\KwOut{Encrypted attack type $ y_t^{\text{cloud}} $.}
Obtain the encrypted graph embedding sequence $ \{\textit{\textbf{g}}_{t-T+1}, \dots, \textit{\textbf{g}}_t\} $ using GAT\;
Process the encrypted sequence through LSTM to compute $ \textit{\textbf{s}}_T $\;
Classify $ y_t^{\text{cloud}} \leftarrow \arg\max \left( \text{softmax}(\textit{\textbf{W}}_{\text{ADU}} \textit{\textbf{s}}_T + \textit{\textbf{b}}_{\text{ADU}}) \right) $\;
\Return $ y_t^{\text{cloud}} $\;
\end{algorithm}

\subsection{Training}
The ETU’s binary classifier is trained offline using labeled normal and attack data. We first pretrain the GAT and LSTM modules separately on historical datasets, then fine‑tune the event-triggered GAT-LSTM combined framework end‑to‑end to optimize both local detection accuracy and cloud‑side classification performance.

\section{Numerical Results}

In this section, we evaluate the proposed attack detection framework using datasets from the LBNL HVAC Fault Detection and Diagnostics (FDD) repository \cite{lbnl_fdd}. These datasets, originally designed for fault detection in rooftop units (RTUs), are repurposed to simulate various attack scenarios on HVAC systems. We describe the dataset, experimental setup, and comparison algorithms used in our study, followed by an analysis of the results.

\subsection{Dataset Description}

\begin{figure}[h]
    \centering
    \includegraphics[width=2.5in]{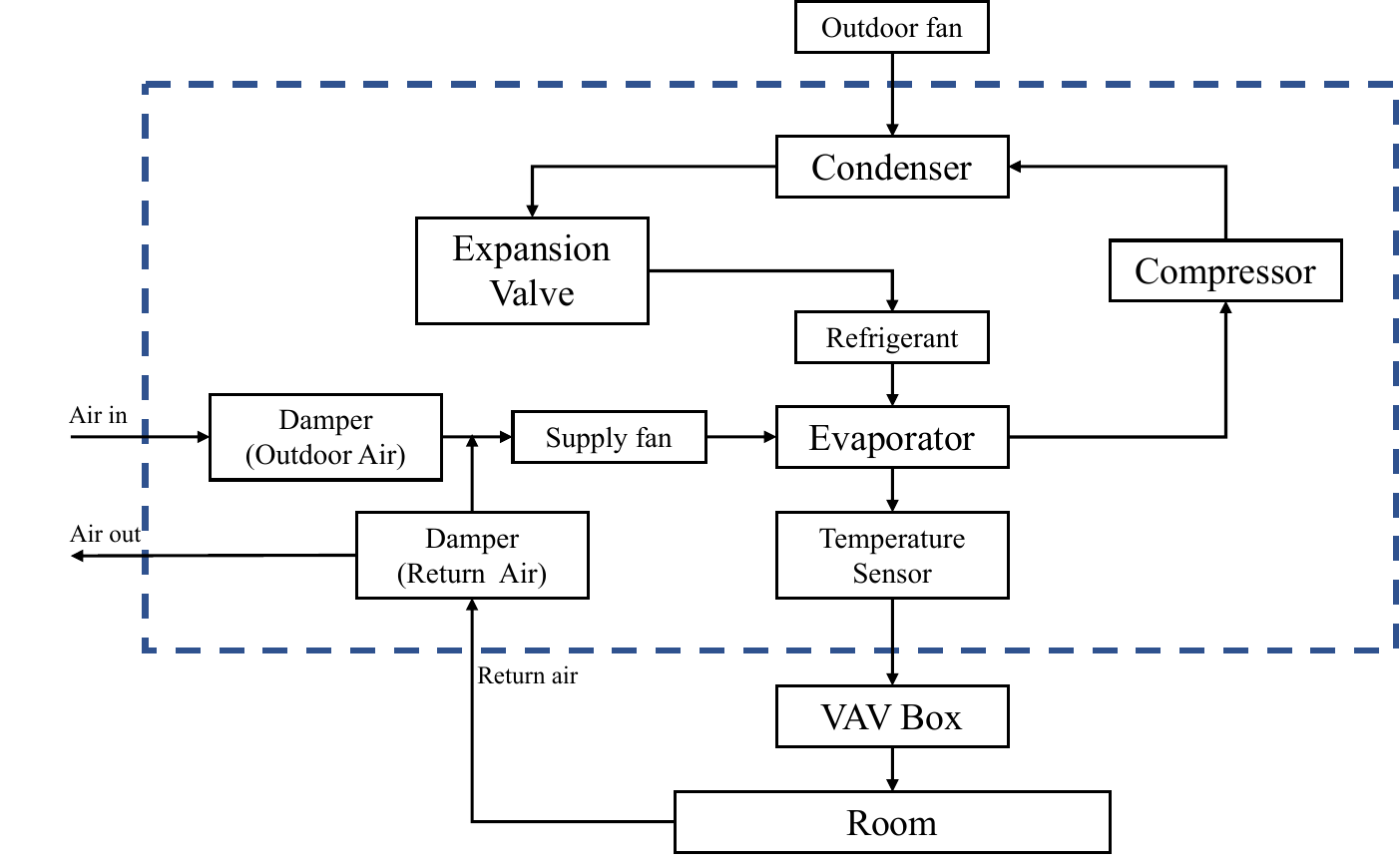}
    \caption{Schematic diagram of experimental RTU.}
    \label{fig:rtu}
\end{figure}

The datasets used in this study are derived from the LBNL HVAC FDD repository. These datasets have been adapted to simulate various attack scenarios on HVAC systems. The RTU, a commonly used component in commercial HVAC systems, consists of interconnected elements such as compressors, evaporators, condensers, economizers, and sensors, which collaboratively maintain desired environmental conditions by regulating airflow, temperature, and CO$_2$ concentration. A schematic diagram of the system is shown in Fig. \ref{fig:rtu}. Note that in each component in this diagram, there are data points that are used in training.

The dataset includes measurements collected under both normal and anomalous conditions. In this study, we reinterpret those anomalies as attack scenarios rather than simple faults. Specifically, five types of attack scenarios are considered: refrigerant undercharge or overcharge, evaporator coil fouling, condenser coil fouling, economizer faults (\emph{e.g.}, stuck or biased dampers), and temperature sensor bias.
Each of these attack types represents potential disruptions caused by adversarial manipulations of the HVAC system, either through physical interventions or cyber attacks on sensor data.

The dataset contains 24 features, including supply and return air temperatures, damper positions, compressor states, \emph{etc}. For graph representation, each RTU component is treated as a node. For example, as RTU components shown in Fig. \ref{fig:rtu}, the condenser and compressor are regarded as nodes.  The physical or logical connections between components are modeled as edges. The graph structure enables the framework to capture both local interactions and global system dynamics effectively.

\subsection{Experimental Setup}
To evaluate the proposed framework, the dataset is preprocessed and represented as a graph-based structure. 
Each graph corresponds to the HVAC system where nodes represent RTU components and edges encode their physical connections. We construct one graph for each time step in our dataset. We split the dataset into 80\% training and 20\% testing samples, ensuring a balanced representation of normal and anomalous states. The ETU filters out normal sequences to reduce data transmission, while the ADU classifies suspicious sequences based on 7 historical data. This window size was chosen by experiment to capture temporal dependencies while keeping computational demands reasonable. The GAT component uses two layers, with four attention heads in the first layer and one in the second. The LSTM layer uses 64 hidden units to model sequential dynamics.

We evaluate the proposed framework using standard classification metrics—accuracy, precision, recall, and F1-score.


\subsection{Baseline Algorithms}
To evaluate the performance of our proposed framework, we compare the ADU component, which performs attack classification, with several baseline methods. In these experiments, the ETU remains unchanged, pre-selecting potential anomalies for transmission, while the ADU is replaced by alternative methods. We also assess the encryption compatibility of each baseline method. To this end, we define 
\[
\textit{\textbf{x}}_t = \begin{bmatrix} x_t(1) \\ x_t(2) \\ \vdots \\ x_t(m) \end{bmatrix} \in \mathbb{R}^m
\]
as the measurement vector at time $t$, where $x_t(i)$ is its $i$-th feature and $m$ is the total number of features.

\paragraph{Threshold-Based Detection}  
For each feature $x_t(i)$, the anomaly score is computed as
\begin{equation}
     S_t(i) = \left| x_t(i) - \mu{(i)} \right|,
\end{equation}
where $\mu{(i)}$ is the expected mean of the $i$-th feature under normal conditions. An anomaly is flagged if $S_t(i)> \tau$ (with $\tau$ as a predefined threshold). This method is adaptable to encrypted data via Order-Preserving Encryption (OPE), because this method needs to compare the number under encryption, OPE maintains the order of values.

\paragraph{k-Nearest Neighbors (kNN)}  
kNN \cite{knn} classifies a test vector $\textit{\textbf{x}}_t$ by comparing it to a training set $\{\textit{\textbf{x}}_j\}_{j=1}^N$, where each $\textit{\textbf{x}}_j \in \mathbb{R}^m$ is a measurement vector. The Euclidean distance between $\textit{\textbf{x}}_t$ and $\textit{\textbf{x}}_j$ is defined as
\begin{equation}
    d\big(\textit{\textbf{x}}_t, \textit{\textbf{x}}_j\big) = \sqrt{\sum_{i=1}^{m} \left( x_t(i) - x_j(i) \right)^2},
\end{equation}
with $x_j(i)$ denoting the $i$-th feature of $\textit{\textbf{x}}_j$. The predicted label is determined by majority voting among the $k$ nearest neighbors, and the simplicity of this computation allows adaptation to encrypted data.

\paragraph{Support Vector Machine (SVM)}  
SVM \cite{svm} determines a decision boundary via the function
\begin{equation}
    y_t = \text{sign}\Big(\textit{\textbf{w}}^\top \textit{\textbf{x}}_t + \textit{\textbf{b}}\Big),
\end{equation}
where $\textit{\textbf{w}} \in \mathbb{R}^m$ is the weight vector and $\textit{\textbf{b}} \in \mathbb{R}$ is the bias term. Since SVM relies on dot products and additions, it is compatible with encryption.

\paragraph{Convolutional Neural Networks (CNN)}  
CNNs \cite{cnnreview} extract temporal features from a sequence
\[
\{\textit{\textbf{x}}_{t-L+1},\, \dots,\, \textit{\textbf{x}}_{t-1},\, \textit{\textbf{x}}_{t}\}
\]
over a window of length $L$. For a given filter $\theta$, the convolution operation is defined as
\begin{equation}
    \textit{\textbf{h}}_\theta = \sigma\Bigg( \sum_{\ell=1}^{L} \textit{\textbf{w}}_\theta[\ell] \cdot \textit{\textbf{x}}_{t-\ell+1} + \textit{\textbf{b}}_k \Bigg),
\end{equation}
where $\textit{\textbf{w}}_\theta[\ell]$ are the filter weights, $\textit{\textbf{b}}_\theta$ is the bias, and $\sigma(\cdot)$ is a non-linear activation function. Although effective, CNNs are computationally intensive when applied to encrypted data.

\paragraph{Fully Connected Neural Networks (FCNN)}  
FCNNs or MLPs model non-linear relationships via layers computed as
\begin{equation}
    \textit{\textbf{h}}^{l} = \sigma\Big( \textit{\textbf{W}}^{l} \textit{\textbf{h}}^{l-1} + \textit{\textbf{b}}^{l} \Big), \quad \textit{\textbf{h}}^{0} = \textit{\textbf{x}}_t,
\end{equation}
where $\textit{\textbf{W}}^{l}$ and $\textit{\textbf{b}}^{(l)}$ are the weight matrix and bias vector for layer $l$. Despite their flexibility, FCNNs lead to high computational costs when used with encrypted data.

\paragraph{Compatibility with Encryption}
Table~\ref{tab:encryption} summarizes the encryption compatibility of the comparison algorithms. Methods such as kNN and SVM, which rely on simple mathematical operations, are more suitable for encrypted processing. In contrast, deep learning models (\emph{e.g.}, CNN and FCNN) face significant challenges due to the high computational cost of operations on encrypted data.


\begin{table*}[ht]
\centering
\caption{Encryption Compatibility of Comparison Algorithms}
\label{tab:encryption}
\begin{tabular}{lcc}
\hline
\textbf{Algorithm} & \textbf{Supports Encryption} & \textbf{Notes} \\
\hline
Threshold-Based     & Yes & Compatible with OPE for comparisons. \\
kNN & Yes & Distance-based operations support encryption. \\
SVM                 & Yes & Linear SVMs are encryption-friendly. \\
CNN                 & Limited & High computational cost for convolutional operations. \\
FCNN                & Limited & Complex operations hinder encryption feasibility. \\
\hline
\end{tabular}
\end{table*}

\subsection{Results and Discussion}

\subsubsection{Performance Comparison}
We evaluate the proposed framework against traditional machine learning models. Table~\ref{tab:results} presents the detection performance of different approaches.

\begin{table*}[ht]
\centering
\caption{Performance Comparison of Attack Detection Methods}
\label{tab:results}
\begin{tabular}{lcccc}
\hline
\textbf{Method} & \textbf{Accuracy (\%)} & \textbf{Precision (\%)} & \textbf{Recall (\%)} & \textbf{F1-Score (\%)} \\
\hline
Threshold-Based       & 72.5 & 70.8 & 74.3 & 72.5 \\
kNN   & 85.2 & 84.7 & 85.9 & 85.3 \\
CNN                   & 93.8 & 93.2 & 94.0 & 93.6 \\
FCNN                  & 90.6 & 89.9 & 91.2 & 90.5 \\ 
Proposed Framework    & \textbf{98.8} & \textbf{98.5} & \textbf{99.0} & \textbf{98.7} \\
\hline
\end{tabular}
\end{table*}

The results indicate that the proposed framework outperforms all baseline models, achieving an accuracy of 98 8\%, demonstrating its effectiveness in detecting cyber-physical attacks in HVAC systems. Compared to deep learning models such as CNN and FCNN, the proposed method provides better classification accuracy, recall, and F1-score, highlighting the advantages of integrating spatial-temporal modeling with an event-triggered mechanism.

\subsubsection{Model Analysis and Communication Efficiency}
To analyze the contribution of each component in the proposed model, we conducted an ablation study testing three variations: one that uses only GAT, another that uses only LSTM and a version without the event-triggering mechanism (ETU). The results in Table~\ref{tab:ablation} show the impact of these modifications.

\begin{table*}[ht]
    \centering
    \caption{Ablation Study and Communication Overhead Results}
    \label{tab:ablation}
    \begin{tabular}{lccc}
        \hline
        \textbf{Model Variant} & \textbf{Accuracy (\%)} & \textbf{F1-Score (\%)} & \textbf{Data Transmission (\%)} \\
        \hline
        GAT-only      & 94.2  & 93.8  & 100  \\
        LSTM-only     & 91.5  & 90.9  & 100  \\
        No ETU        & \textbf{99.0}  & \textbf{98.8}  & 100  \\
        Full Model (Proposed framework) & 98.8 & 98.7 & \textbf{15}  \\
        \hline
    \end{tabular}
\end{table*}

The results highlight that GAT alone achieves higher accuracy than LSTM alone, indicating that capturing spatial dependencies is crucial for attack detection in HVAC systems. The No ETU model, which transmits all data continuously to the cloud, achieves the highest accuracy and F1-score since it avoids local misclassification by the ETU. However, this comes at the cost of 100\% data transmission, making it impractical for real-world deployment. 
The full model achieves a slightly lower accuracy than No ETU, but it reduces data transmission by 85\% while maintaining high classification performance. This tradeoff demonstrates that ETU effectively minimizes network overhead without significantly compromising accuracy, making it a scalable and efficient solution for HVAC systems.

\begin{figure}[t]
    \centering
    \includegraphics[width=3.2in]{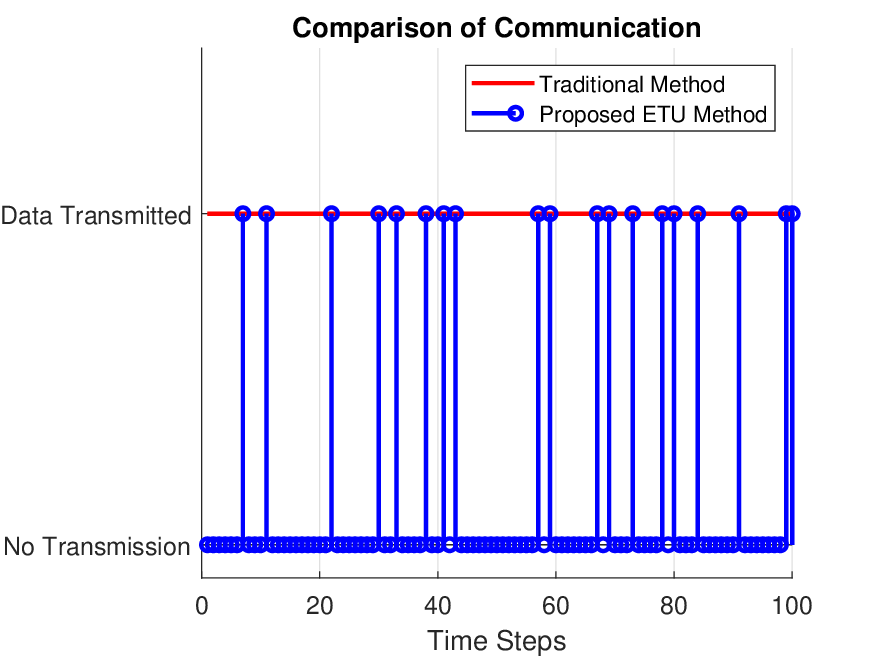}
    \caption{Comparison of communication overhead between the traditional method and the proposed ETU-based framework.}
    \label{fig:comm_overhead}
\end{figure}

As illustrated in Fig.~\ref{fig:comm_overhead}, traditional methods transmit data at every time step, leading to constant communication overhead. In contrast, the proposed ETU-based approach selectively transmits data, resulting in an 85\% reduction in communication overhead. This selective transmission strategy significantly optimizes resource usage, reducing unnecessary cloud computation and network costs, making it well-suited for real-world deployment in resource-constrained environments.

\subsubsection{Classification Performance Analysis}
\begin{figure}[h!]
    \centering
    \includegraphics[width=2.8in]{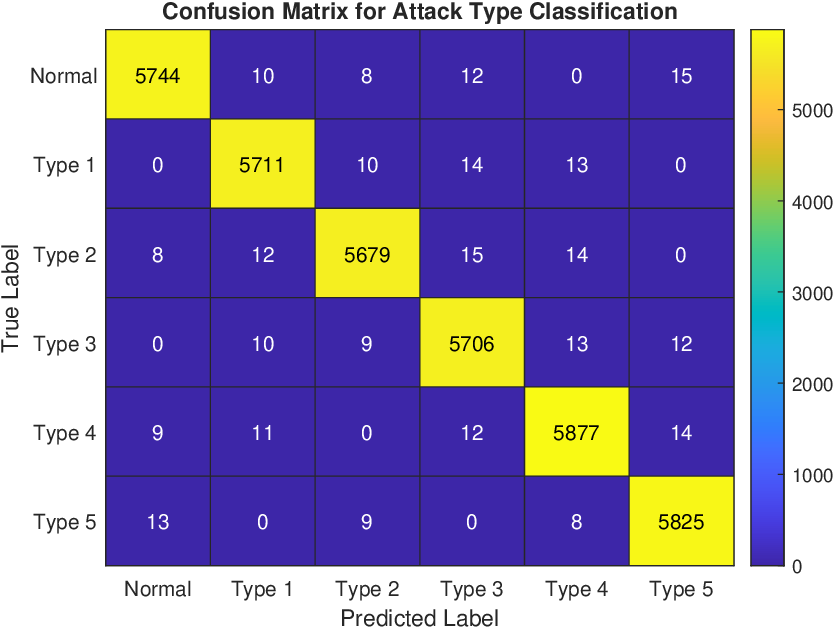}
    \caption{Confusion matrix for attack type classification.} 
    \label{fig:confusion_matrix}
\end{figure}

To further evaluate the attack detection performance, we analyze the classification results using the confusion matrix shown in Fig.~\ref{fig:confusion_matrix}.

The matrix exhibits strong diagonal dominance, which indicates that the vast majority of samples are correctly classified. For instance, the detection accuracy for Type 2 and Type 5 attacks exceeds 97\%, while misclassification rates for other types, such as Type 3, remain below 2\%. This high level of accuracy confirms that the integration of GAT for spatial dependency modeling and LSTM for capturing sequential patterns effectively distinguishes between normal and anomalous states. Although there are a few off-diagonal entries suggesting minor confusions between similar attack patterns, their low frequency demonstrates that the framework maintains robust performance with minimal false positives and false negatives. These results underscore the reliability of the proposed framework for practical deployment in HVAC systems.

\subsection{Simulation of Event‐Triggered GAT-LSTM Detection Workflow}
\begin{figure}[h]
  \centering
  \includegraphics[width=\linewidth]{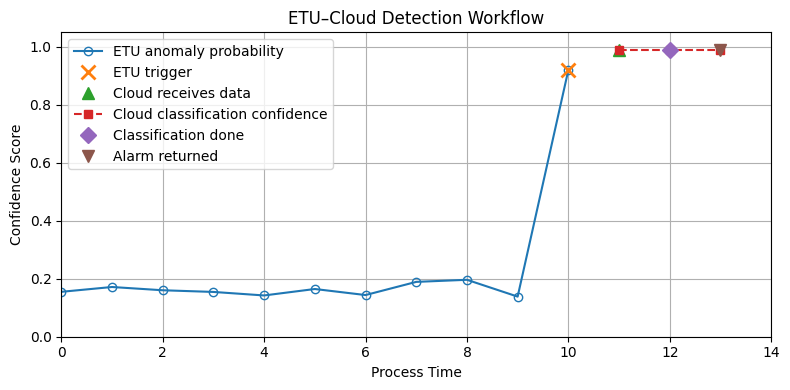}
  \caption{ETU–Cloud detection workflow: ETU anomaly probability (blue $\circ$), ETU trigger (orange \texttimes), cloud receives data (green $\triangle$), cloud classification confidence (red $\square$), classification done (purple $\diamond$), and alarm returned (brown $\triangledown$).}
  \label{fig:workflow}
\end{figure}

To illustrate a single detection cycle, Fig.~\ref{fig:workflow} plots the anomaly probability given by ETU and the cloud’s attack–type confidence against “process time” (steps of the pipeline). The vertical axis shows confidence scores (0–1). Key events are highlighted with distinct markers.

During steps 0–9 the ETU remains in normal mode with low anomaly probabilities (blue $\circ$). At step 10 it crosses its alarm threshold (orange \texttimes), triggering data upload. One step later (step 11, green $\triangle$) the cloud receives the encrypted measurements. The GAT-LSTM classifier then immediately evaluates them. Its high confidence (red $\square$, 0.99) first appears at step 12 (purple $\diamond$ “classification done”), and the alarm is returned at step 13 (brown $\triangledown$ “alarm returned”).  

This timeline emphasizes the sparsity of ETU triggers among many normal samples and the consistently high confidence of the cloud‐side GAT-LSTM classifier.

\section{Conclusion}
This paper presented a novel framework for detecting and classifying cyber-physical attacks in HVAC systems, combining GAT for spatial modeling and LSTM networks for temporal modeling. By introducing an ETU, this framework minimizes communication overhead by transmitting encrypted data to the cloud only when a potential attack is detected.

The proposed method achieves high accuracy and low communication cost compared to traditional methods, as demonstrated by the experiment section. It effectively distinguishes between normal operations and various attack types with minimal misclassifications, ensuring reliable performance in diverse scenarios.

Future work will focus on adaptive learning mechanisms to address evolving and subtle attack patterns, ensuring the framework remains robust and effective for long-term deployment.


\bibliographystyle{IEEEtran}

\bibliography{references.bib}



\vspace{11pt}

\vspace{11pt}

\vfill

\end{document}